\documentclass[aps,showpacs]{revtex4}

\usepackage{epsfig}

\newcommand{\be}{\begin{equation}}
\newcommand{\ee}{\end{equation}}
\newcommand{\bc}{\begin{center}}
\newcommand{\ec}{\end{center}}
\newcommand{\bea}{\begin{eqnarray}}
\newcommand{\eea}{\end{eqnarray}}
\newcommand{\ba}{\begin{array}}
\newcommand{\ea}{\end{array}}

\newcommand{\m}{\mathcal{M}}
\newcommand{\cc}{\mathcal{C}}

\begin{document}

\title{Induced cultural globalization by an external vector field in an enhanced Axelrod model}
\author{Arezky H. Rodr\'{\i}guez$^{1}$, M. del Castillo-Mussot$^{2}$ and G. J. V\'azquez$^{2}$}

\affiliation{$^{1}$ Academia de Matem\'aticas, Universidad Aut\'onoma de la Ciudad de M\'exico,
Mexico City, Mexico. \\
$^{2}$ Departamento de Estado Sólido, Instituto de F\'{\i}sica, Universidad Nacional Aut\'onoma de M\'exico (UNAM), Apdo. Postal 20-364, San \'Angel 01000, M\'exico D.F., Mexico.}

\date{\today}

\begin{abstract}
A new model is proposed, in the context of Axelrod's model for the study of cultural dissemination,
to include and external vector field (VF) which describes the effects of mass media on social
systems. The VF acts over the whole system and it is characterized by two parameters: a non-null
overlap with each agent in the society and a confidence value of its information. Beyond a
threshold value of the confidence there is induced monocultural globalization of the system lined
up with the VF. Below this value, the multicultural states are unstable and certain homogenization
of the system is obtained in opposite line up according to that we have called {\it negative
publicity} effect. Three regimes of behavior for the spread process of the VF information as a
function of time are reported.
\end{abstract}



\maketitle

\section{Introduction.}

Agent-Based Models (ABMs) \cite{bonabeau,robert2005} are computer si\-mu\-la\-tions of the local
interactions of the members of a population which could be plants and animals in ecosystems
\cite{grimm}, vehicles in traffic, people in society \cite{axelrod-original-paper},
etc. Locally interactions at lower-level give rise to the spontaneously emergence of
higher-level organizations whose properties are not possessed by the individuals neither directly
determined by them. Complex and non-linear phenomena have attracted the attention of the
scientific community to study the interplay between the lower and higher levels of organizations
\cite{Laughlin-libro}. These models typically consist of an environment or framework in which the
interactions occur among some number of individuals defined in terms of their behaviors (procedural
rules) allowing the tracking of the characteristics of each individual through time.

There are lots of applications to model different aspects of dynamics in society
\cite{sociodynamics,quantitative-sociodynamics,castellano}. Specifically, the Axelrod model
\cite{axelrod-original-paper,axelrod-libro,vespignani} is an ABM designed to investigate the
dissemination of culture among interacting agents in a society. In this model, society is
represented by a lattice composed by a 2-dimensional array of vectors (agents) with a number of
entries called ``features''. The definition of its culture is given by the set of traits an agent has in its features. The Axelrod's model has been exhaustively implemented to study a great variety of
problems: the nonequilibrium phase transition between monocultural and multicultural states
\cite{klemm2003-2}, the cultural drift driven by noise \cite{klemm2003-1,sanctis}, nominal and
metric features \cite{flache,jacobmeier}, propaganda \cite{carletti2006}, time evolution dynamics
\cite{vazquez2007}, the resistance of a society to the spread of a foreign cultural traits
\cite{boccara}, finite size effects \cite{toral2006}, the impact of the evolution of the network
structure with cultural interaction \cite{centola2007} among others.

\begin{figure}
\vspace{3cm}
\psfig{file=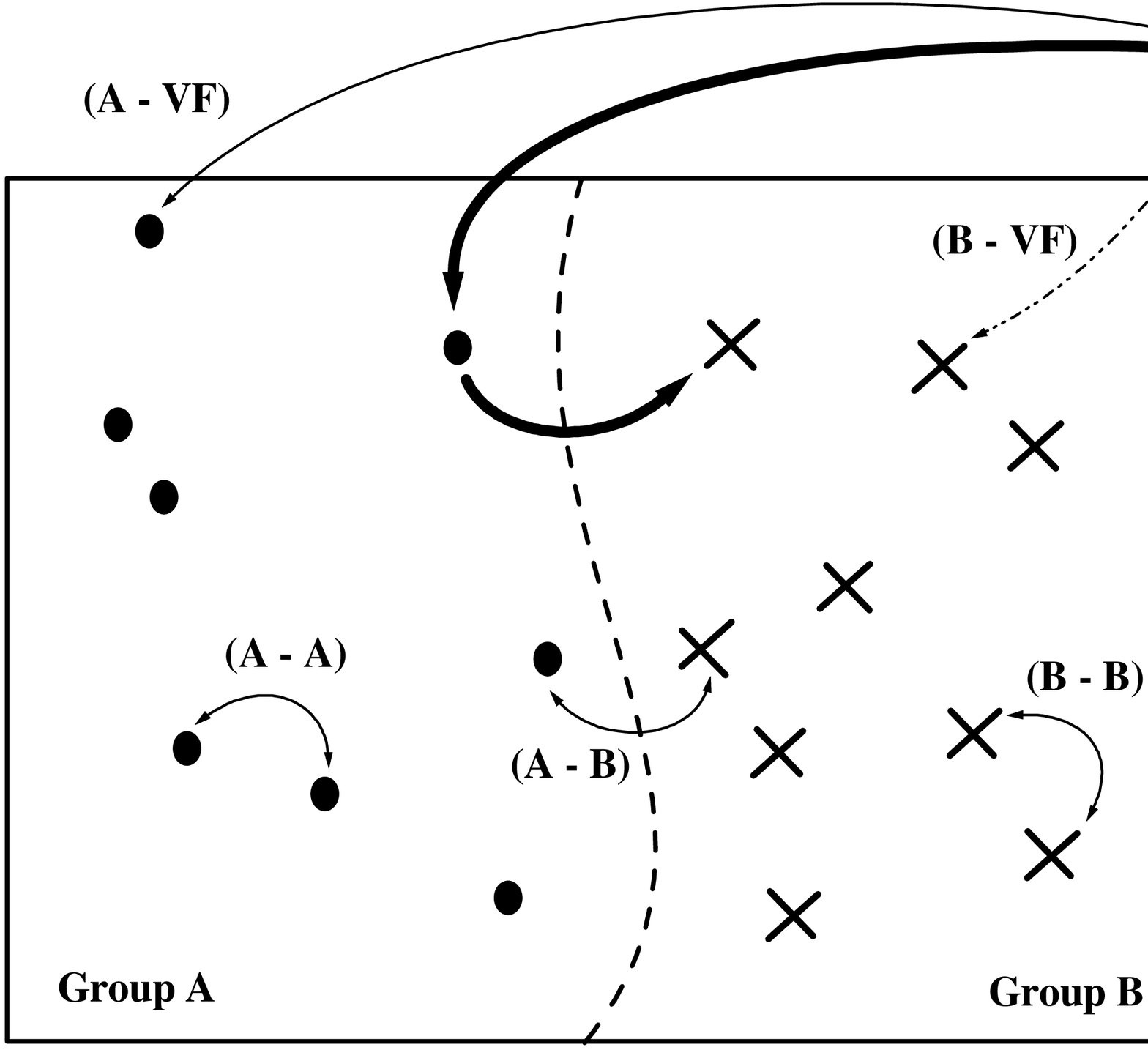,width=90mm}
\vspace{-3cm}
\caption{Former models representation of agents and dynamics interactions in the lattice when the
VF is included (represented by a big square). Agents from group A are indicated with dots while
agents in group B are indicated with crosses. Different interactions are represented with different
lines. The diffusion mechanism from the VF to agents in group B is represented by thick lines.
Null direct interaction between VF and agents from group B is represented by dash-dot-dot line.}
\label{fig0}
\end{figure}

Some works have been done including in the Axelrod model an extra agent acting as a vector
field (VF) over the whole society with the purpose of simulating a mass media effects
\cite{shibanai2001,avella2005,avella2006,avella2007}. In all of them, the interaction between the
external VF and agents is similar to that between an agent and its neighbors: they interact only
if they have at least one common trait in their corresponding features. In this formalism, the inclusion
of an external field, which does not change its values on time, introduces an asymmetry on the
lattice, which can now be described as composed by two groups of agents: group A where agents have
trait(s) in common with the VF and group B whose do not. All the interactions can now be classified as
follows: agents from group A with VF (A-VF), agents from group B with VF (B-VF), between agents
from group A (A-A), between agents from group B (B-B) and finally between agents from group A and B
(A-B). See Fig. \ref{fig0} and lines therein. The (B-VF) interaction, showed as dash-dot-dot line
in the figure, is a null interaction because agents from group B do not share traits with the VF.
In this way, the only opportunity of agent B to acquire one VF trait is through a diffusion
mechanism with the combined interactions (A-VF) plus (A-B) as pointed out in Fig. \ref{fig0} with a
thick line. In those models it is also included the strength of the VF as a probability $P$ of
interactions A-VF and B-VF while the probability of interactions A-A, B-B and A-B are given then by
$1-P$. Therefore the diffusion mechanism, showed with a thick lines in Fig. \ref{fig0}, has very low
probability as $P$ increases and agents from group B are set apart from the VF information.
Furthermore, the mechanism B-VF can be very active (time consuming) but with null effects, and
the internal relaxing mechanism B-B is not able to drive agents to the final state in an efficient
way. Thus, the final absorbing states obtained in those previous models consisting in multicultural states when the VF strength is increased is then not surprising \cite{shibanai2001,avella2005,avella2006,avella2007}.

Our current interest is to develop a new model for the inclusion of an extra agent acting as a
vector field (VF) over the {\it whole} society to overcome the difficulties achieved by the previous
models described above. As mentioned in Ref. \cite{shibanai2001}, the media information is socially
processed through personal networks. Then, models will be more realistic is they allow a strong interaction with the VF without loss of interchanges between agents on the lattice.
It is also important to say that mass media designs its publicity in a clever way. As mentioned by the
anthropologist Gregory Bateson: to produce a change it is necessary to be different but, at
the same time, it is necessary to be ``close enough'' to be taken into account \cite{bateson}. When
acting over the society, mass media always try to have something in common with the people chosen
as target of publicity or propaganda. It is designed to offers attractive materials for the whole
society: news, sports, soap operas, movies, cartoons, music, arts, etc.

Our goal here is to develop a model to include this effect considering an additional non-zero
probability of all agents to copy a trait from the VF, even if they do not share any trait of their
features. Section \ref{the-model} is devoted to that purpose. In section \ref{num-results} it is
exposed some numerical calculations and finally some conclusions are outlined in section
\ref{conclusiones}.

\section{\label{the-model} The model.}

The system consists of $L^2$ agents as the sites of a square lattice. The state of an agent $i$ is
defined as a vector of $F$ {\it nominal} components called features given by $\sigma_i =
(\sigma_{i_1}, ..., \sigma_{i_f}, ..., \sigma_{i_F})$ which characterize the nominal
$F$-dimensional culture of the corresponding agent. This way, each agent has four nearest neighbors
but as the fifth it is introduced the VF $\m$ with nominal features $\sigma_\m = (\sigma_{\m_1},
..., \sigma_{\m_f}, ..., \sigma_{\m_F})$. The VF intents to simulate and external mass media or
publicity which acts over the whole society. Then, each agent can interact with five agents: its
four nearest neighbors and the VF $\m$, all with equal probability $1/5$. Additionally, each feature
$\sigma_{i_f}$ and $\sigma_{\m_f}$ can take any of the values in the set $\{0,1, ..., q - 1\}$
which are the corresponding cultural traits of an agent $i$ or the VF $\m$. Initially, the values
of the vectors $\sigma_i$ and $\sigma_\m$ are randomly and independently set with one of the $q^F$
state vectors with uniform probability.

\begin{figure}
\vspace{2cm}
\psfig{file=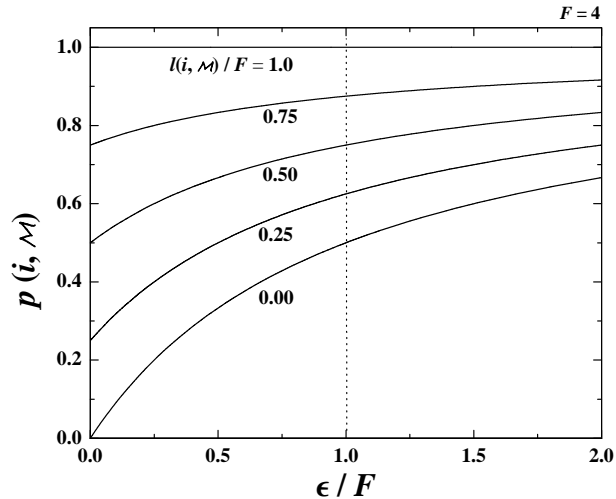,width=100mm}
\vspace{-2.5cm}
\caption{Probability of interaction between the agent $i$ and the vector field $\m$ as a
function of $\epsilon/F$ for fourth nominal traits ($F$=4) and different values of the overlap
$l(i,\m)$.}
\label{fig1}
\end{figure}

The interaction between different agents is possible only when the two vector have an overlap $0 <
l < 1$ where the overlap between two agent $i$ and $j$ is the number of shared traits and it is
given by $l(i,j)=\sum_{f=1}^F \delta_{\sigma_{i_f},\sigma_{j_f}}$. Here $\delta$ is the Kronecker
symbol. The probability, which we call here {\it nominal} probability, of the interaction between
two agents is given by $p(i,j)= l(i,j)/F$. In general, the situation $p(i,j)=0$ is possible when the
overlap between two agents is zero, but the case where the probability between an agent and the VF
is zero is not an acceptable situation for a publicity (or mass media) which intentionally designs
its interaction in such a way that always there are features which have traits in common with the
agent subject of the influence to guarantee that the connection is active.

In order to include this important effect in our model, we included some {\it effective} features
that the VF always shares with each agent when interacting besides nominal features with
the purpose to simulate phenomenologically in a simple way the almost omniscient force of today's
publicity or propaganda in mass media that offers something for all tastes and ages (magazines, radio programs, TV series, etc). Note that the effective features are related with the dynamics between the
agents and the VF while the nominal features are related with the dynamics between agents inside
the lattice. The specific nature in real society of the effective features is not of importance
here. It will be different for different agents, but the intention is to take into account the
specific design of the publicity that mass media does to attract everyone. For $\epsilon/F < 1$ where $\epsilon$ is defined as the ``effective feature'', the
VF and the agent share more nominal than effective features and the VF constitutes a
``perturbation'' to the internal interaction between different agents in the society. The case
$\epsilon/F > 1$ means that there are more effective features that certainly share each agent with
the VF than the number of nominal features each agent has. Then it is now the society which can be
considered as a ``perturbation'' with respect to the more robust dynamics between the VF and the
agents. In our model, the parameter $\epsilon$ not only takes natural values, but also fractional
values, as it will be seen later.

Therefore, the probability of interaction between the external vector and an agent, which we call
here {\it extended} probability, is written as
\be
\label{effective-probability}
p(i,\m) = \frac{l(i,\m) + \epsilon}{F + \epsilon}
\ee
or, in dimensionless parameters,
\be
p(i,\m) = \frac{l(i,\m)/F + \epsilon/F}{1 + \epsilon/F}
\ee
where $l(i,\m)$ is the overlap of the nominal features between agent $i$ and the VF.

In Fig. \ref{fig1} are shown the values of the probability $p(i,\m)$ as a function of
$\epsilon/F$ for fourth nominal features ($F=4$) and different values of $l(i,\m)$. The values of
$l/F =$ 0.00, 0.25, 0.50, 0.75 and 1.00 are obtained when the agent shares with the vector field 0,
1, 2, 3 and 4 nominal features. As seen, the probability is zero only when there are not effective
features ($\epsilon=0$) and the overlap between the agent $i$ and the VF is zero. In contrast, in
all the other cases the probability $p(i,\m)$ is always different from zero. For $\epsilon = 0$ the
values for the case with no effective features are recovered. As expected, the probability is lager
for larger values of the effective features $\epsilon$ for a given value of $l(i,\m)$ and also
increases for larger values of the overlap $l(i,\m)$ at a given number of $\epsilon$. Finally, when
the agent $i$ and the VF share all the nominal features ($l(i,\m)=1$) the probability is always one
for any value of $\epsilon$.

\begin{figure}
\vspace{2cm}
\psfig{file=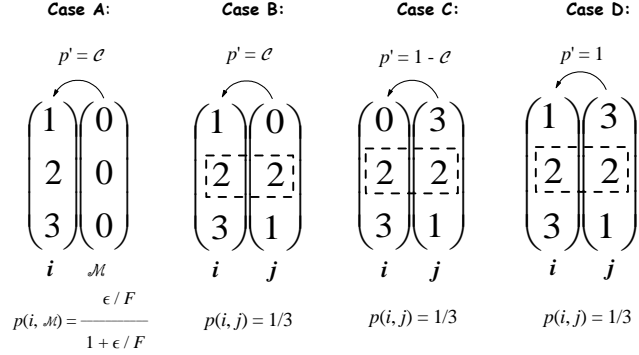,width=90mm}
\vspace{-3.5cm}
\caption{Four possible cases of interaction for a system with $F=3$ features. Shared features are indicated
inside a dashed rectangle. The probability of interaction $p(i,j)$ (or $p(i,\m)$ in Case A) is
indicating below. In each case the trait in $\sigma_{i_1}$ will be deleted by coping trait
$\sigma_{j_1}$ (or by trait $\sigma_{\m_1}$ in Case A). The probability of coping/deleting trait 1
is given by a) $p'=\cc$, b) $p'=\cc$, c) $p'=1 - \cc$ and c) $p'=1$.}
\label{fig2}
\end{figure}

In Ref. \cite{mazzitello2007} the authors have used an expression similar to Eq.
(\ref{effective-probability}), but with $\epsilon = 1$ and, as in Ref. \cite{avella2005}, they have
modeled the strength of the VF as a probability of the interaction between agents and the VF.
Increasing values of this probability implies a decreasing value of the probability for agents
interacting between each other and then the corresponding diffusion of traits values between
agents can be stopped which is not a realistic or desirable effect. They have found that only monocultural states are obtained and only by introducing a noise rate it is possible to drive the system to a multicultural final state.

In our case, we are not interested to include the effects of random perturbation effects, but we instead
introduce another parameter related with the confidence of the information belonging to the VF. As ``confidence'' we understand here the credibility granted by agents to the information possessed by the VF. It
is included as an extra probability $\cc$ for agent $i$ to copy an entry directly from the VF or an entry
from another agent $j$ that belongs to the VF either. It is also included as an extra probability $1 - \cc$ when agent $i$, when coping a trait, deletes and information the VF possesses in the same feature.

To clarify this important concept we show in Fig. \ref{fig2} four situations of interaction which resume all
the possible cases. In Case A it is described an interaction between agent $i$ and the VF which has
been set to (0,0,0) without lost of generality. None of the nominal features are shared and then
the probability of interaction $p(i,\m)$ depends only from the value of the effective features
$\epsilon$ according to the expression in the figure and is always larger than zero. In the
practical case when agent $i$ copies, for example the first entry, then the VF will be copied with
probability $p' = \cc$ which characterize the confidence of the information possesses by the VF. In
the next cases, the interaction occurs between agents $i$ and $j$ which only share one trait of
three possibles. The probability of interaction is then given by $p(i,j) = 1/3$ in all these cases.
In Case B the nominal feature that agent $i$ selects to copy from agent $j$ coincides with the
value the VF has in the same feature. Then, as in Case A, the corresponding trait is copied with
probability $p'=\cc$. In Case C, when coping, agent $i$ will delete its trait which is equal to
that possessed by the VF. Then, it is deleted with probability $p' = 1 - \cc$. Finally, in Case D
the traits copied and deleted are not related with the VF and then they are copied with probability
$p'=1$. Note that according with these rules of interaction, when the confidence of traits
possessed by the VF are $\cc=1$, these traits are always copied with probability $p'=1$ and never
deleted. Otherwise, if the confidence of the traits possessed by the VF is $\cc=0$, this traits are
never copied ($p'=0$) and are always deleted with probability $p'=1$. Then, starting from the
initial condition described above, the system evolves by iterating the following steps:
\begin{enumerate}
\item[(1)] Select at random an agent $i$ on the lattice, which is the active element.
\item[(2)] Select at random, with equal probability, an agent of interaction. It could be one of
the four nearest neighbors or the VF.
\item[(3)] Calculate the overlap $l(i,s)$ where $s=j$ for the neighbor or $s=\m$ for the VF. If $s=\m$,
the agent $i$ and the VF interact with the extended probability $p(i,\m)$. If $s=j$ and $ 0 <
l(i,j) < F$, agents $i$ and $j$ interact with the nominal probability $p(i,j)$.
\item[(4)] In case of interaction between agent $i$ and agent $s$, choose a position trait $h$ at
random such that $\sigma_{i_h} \neq \sigma_{j_h}$ (or $\sigma_{i_h} \neq \sigma_{\m_h}$) and then
set $\sigma_{i_h} = \sigma_{j_h}$ (or $\sigma_{i_h} = \sigma_{\m_h}$) according to:
\begin{enumerate}
\item[(4.1)] if $s=\m$ then set $\sigma_{i_h} = \sigma_{\m_h}$ with probability $p'=\cc$,
\item[(4.2)] if $s=j$ and $\sigma_{j_h}   =  \sigma_{\m_h}$ then set $\sigma_{i_h} = \sigma_{j_h}$
with probability $p'=\cc$,
\item[(4.3)] if $s=j$ and $\sigma_{i_h}   =  \sigma_{\m_h}$ then set $\sigma_{i_h} = \sigma_{j_h}$
with probability $p'=1 - \cc$.
\item[(4.4)] if $s=j$ and both $\sigma_{j_h} \neq \sigma_{\m_h}$ and $\sigma_{i_h} \neq
\sigma_{\m_h}$, then set $\sigma_{i_h} = \sigma_{j_h}$ with probability $p'=1$.
\end{enumerate}
\end{enumerate}

Then, the full probability of agent $i$ copies a trait from agent $j$ is given by
\be
\mathcal{P}(i,j) = \frac{1}{5} \; p(i,j) \; p'
\ee
while the full probability the agent $i$ copies a trait from the VF is given by
\be
\mathcal{P}(i,\m) = \frac{1}{5} \; p(i,\m) \; p'
\ee

Before studying the effects of the VF in our model let us review the original Axelrod's model. The
computational dynamics of this model ends when the system reaches an absorbing state characterized
by either $l(i,j)=0$ or $l(i,j)=F$ for all pairs of closed neighbors ($i,j$). A class of absorbing
state, given by $q^F$ different configurations is called the ``monocultural'' state which
corresponds to the case where $l(i,j)=F$ for all pairs of closed neighbors. In this case, all
agents in the network share the same trait at each feature ($\sigma_{i_h} = \sigma_{j_h}$ for all
($i,j$)) and the dynamics ends. Another class of absorbing state is called ``multicultural'' state
and consist of at least two (or more) homogeneous domains which agents have cultural traits
completely different. This way, two agents belonging to two different domains have zero overlap.
The multicultural state is reached when each agent in the lattice has full of null overlap with all
its neighbors. In these cases, a domain is given by a set of contiguous sites with identical state
vector.

It has been shown that the system reaches monocultural or multicultural states in dependence of lower
($q < q_c$) or higher values ($q > q_c$) of the cultural diversity $q$ \cite{vespignani}. To
characterize the transition it has been considered two different order parameters: the average
fraction of different cultural domains or the average number of agents in the biggest domain
$<S_{max}>$ normalized to the number of lattice elements.

In our case, it will be shown how our model produces a complex pattern of social behavior with
affinities and repulsions to the influence of the VF in dependence of the effective features
$\epsilon$ and the confidence $\cc$ of the information.

\section{\label{num-results} Numerical results.}

\begin{figure*}
\vspace{1.5cm}
\centerline{\psfig{file=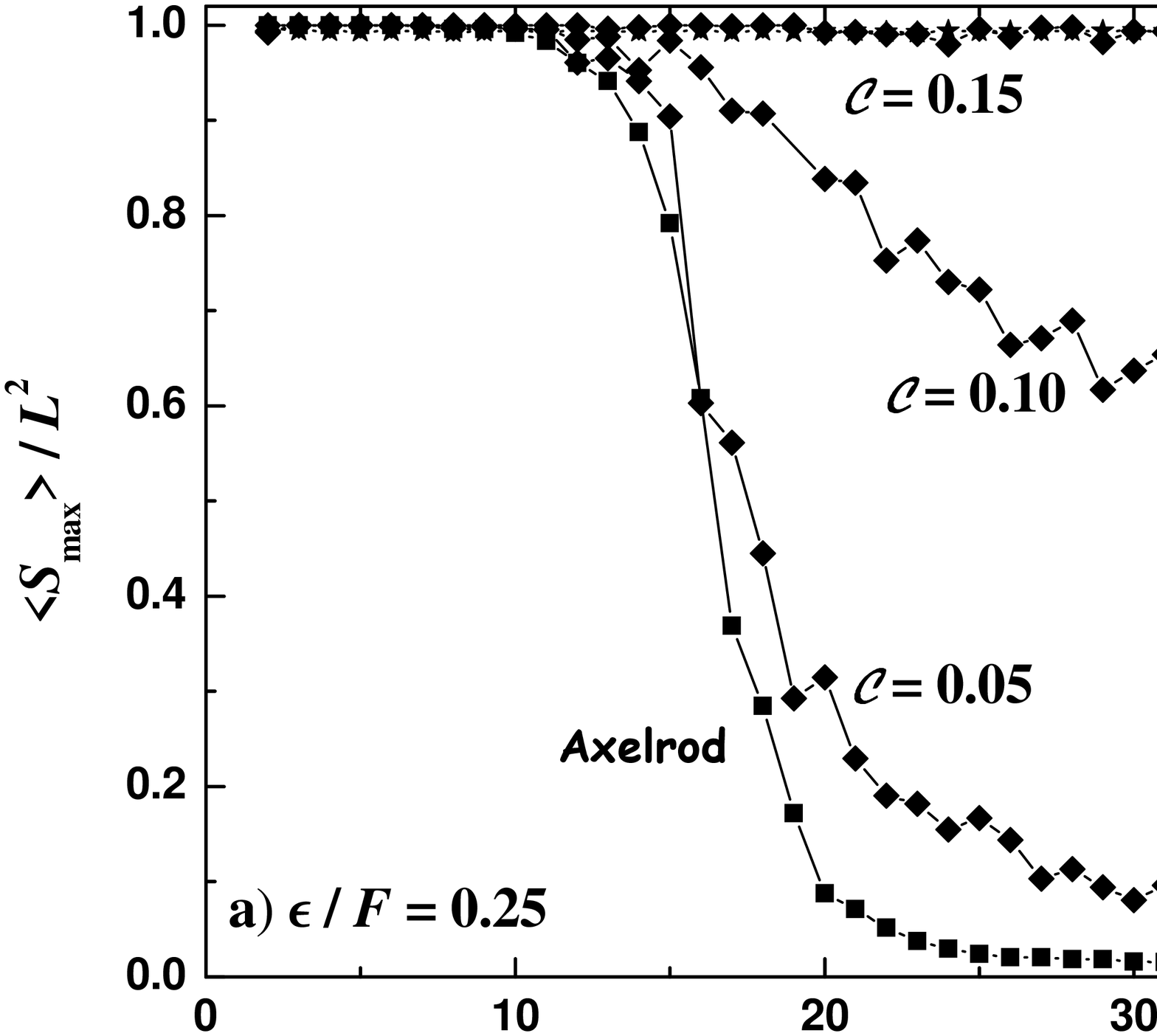,width=80mm} \psfig{file=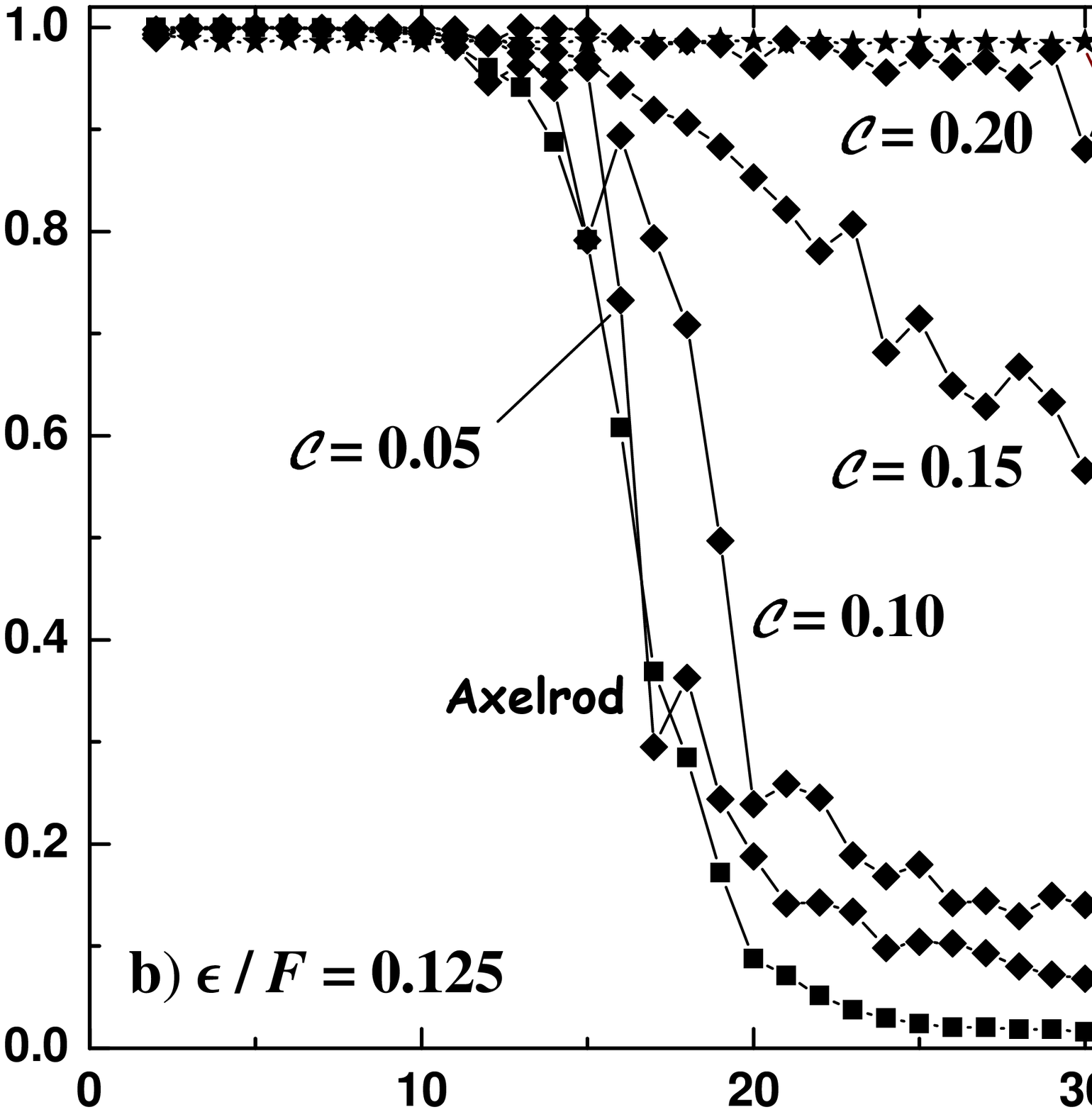,width=80mm}}
\centerline{\psfig{file=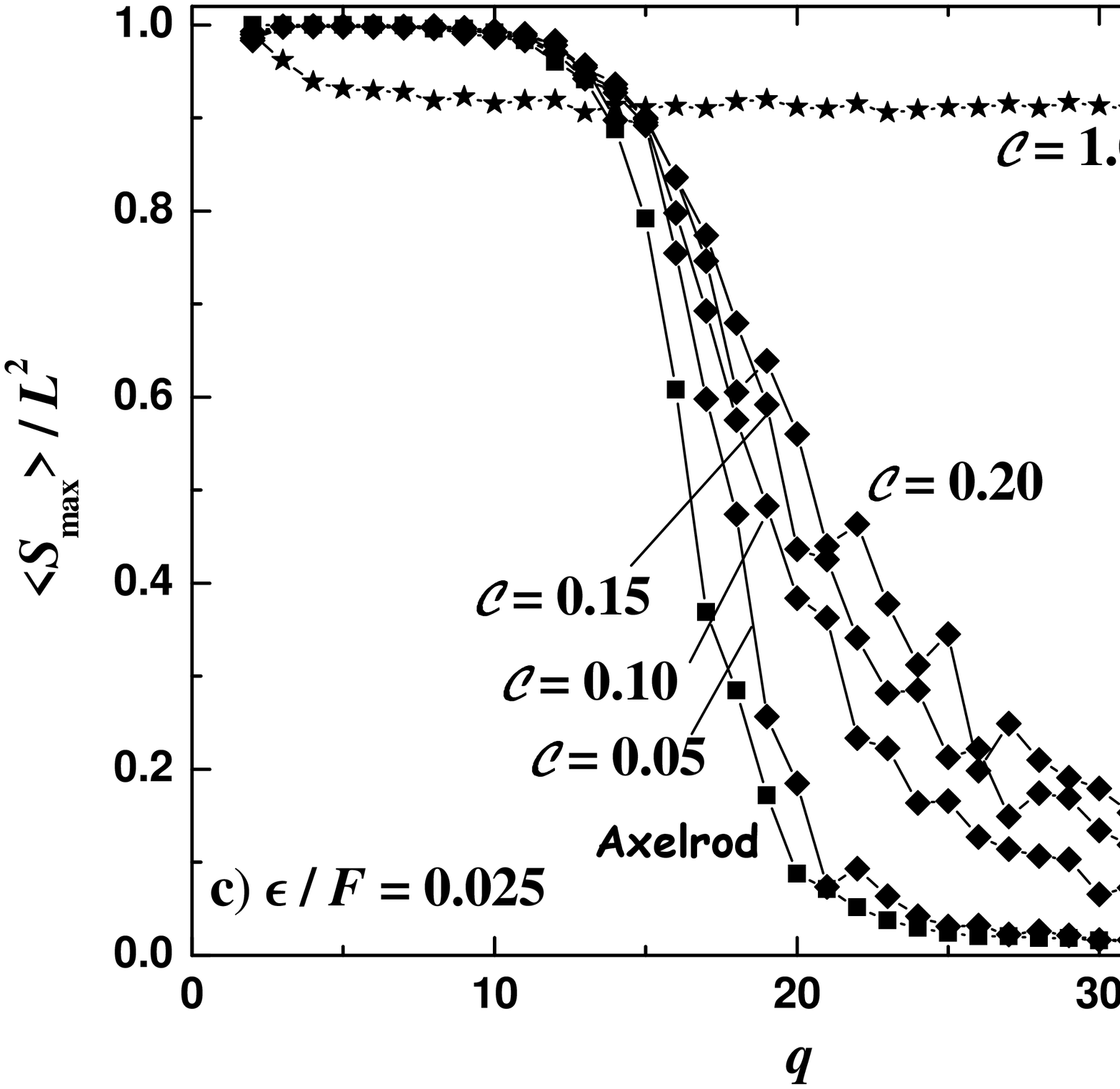,width=80mm} \psfig{file=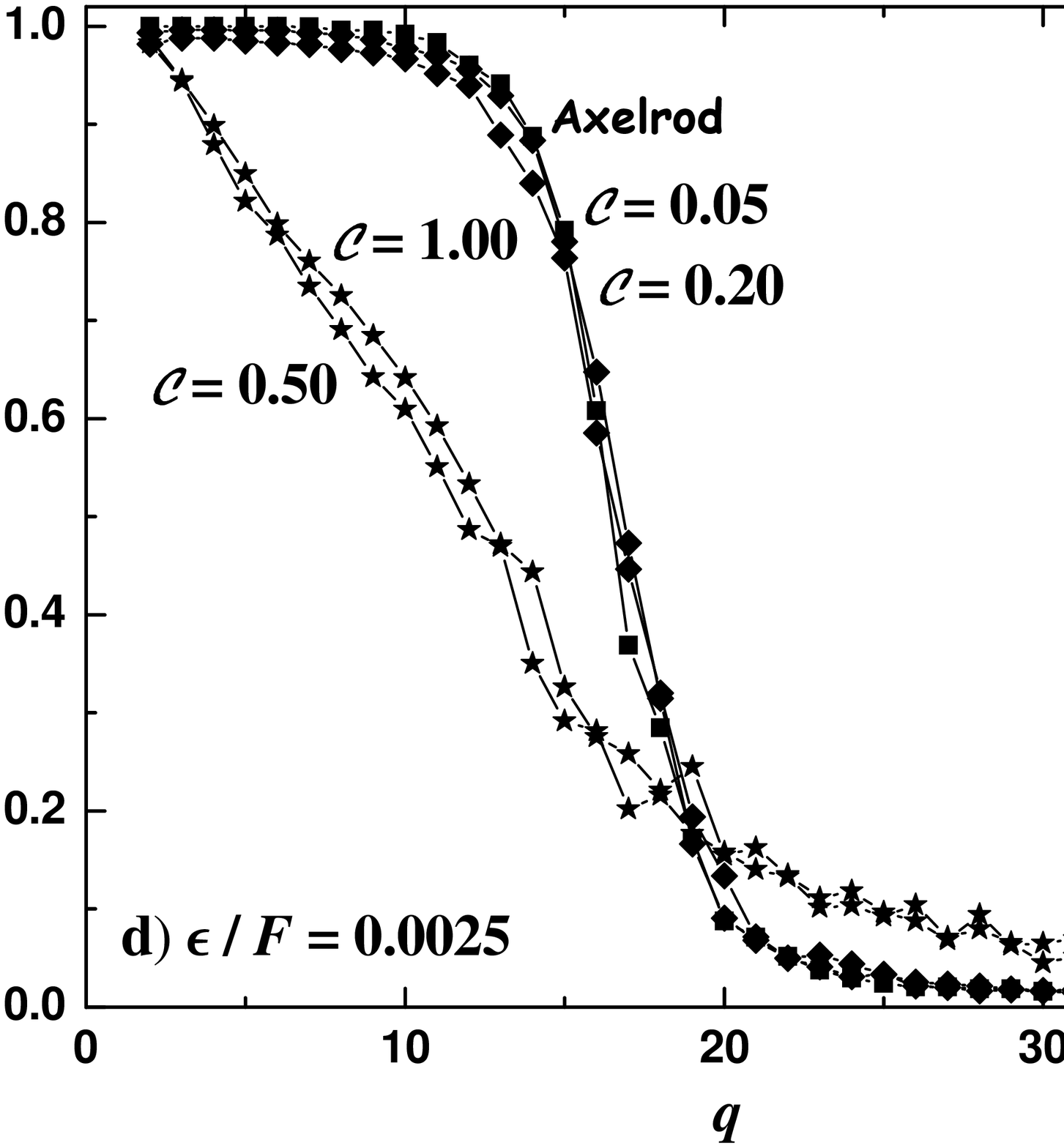,width=80mm}}
\vspace{-2cm}
\caption{Calculation of the normalized average number of agent in the greatest domain
at an absorbing state as a function of $q$ averaged over 50 realizations. The values of the ratio
$\epsilon/F$ are a) 0.25, b) 0.125, c) 0.025 and d) 0.0025. At each panels, different value of the
confidence $\cc$ is taken into account in increasing order. The case of the Axelrod's model without
VF is included in each panel with full square dots.  Full rhombus indicate that the corresponding
absorbing states do not share the information possessed by the external field, while full stars
indicate full coincidence of the absorbing state with the VF.}
\label{fig3}
\end{figure*}

Numerical simulations have been carried out in lattices with $L^2 = 30 \times 30$ agents and $F=4$
features each. Different absorbing states have been found and we report the average
realization number of agent in the largest domain over 50 different initial conditions.

\begin{figure*}
\vspace{2cm}
\centerline{\psfig{file=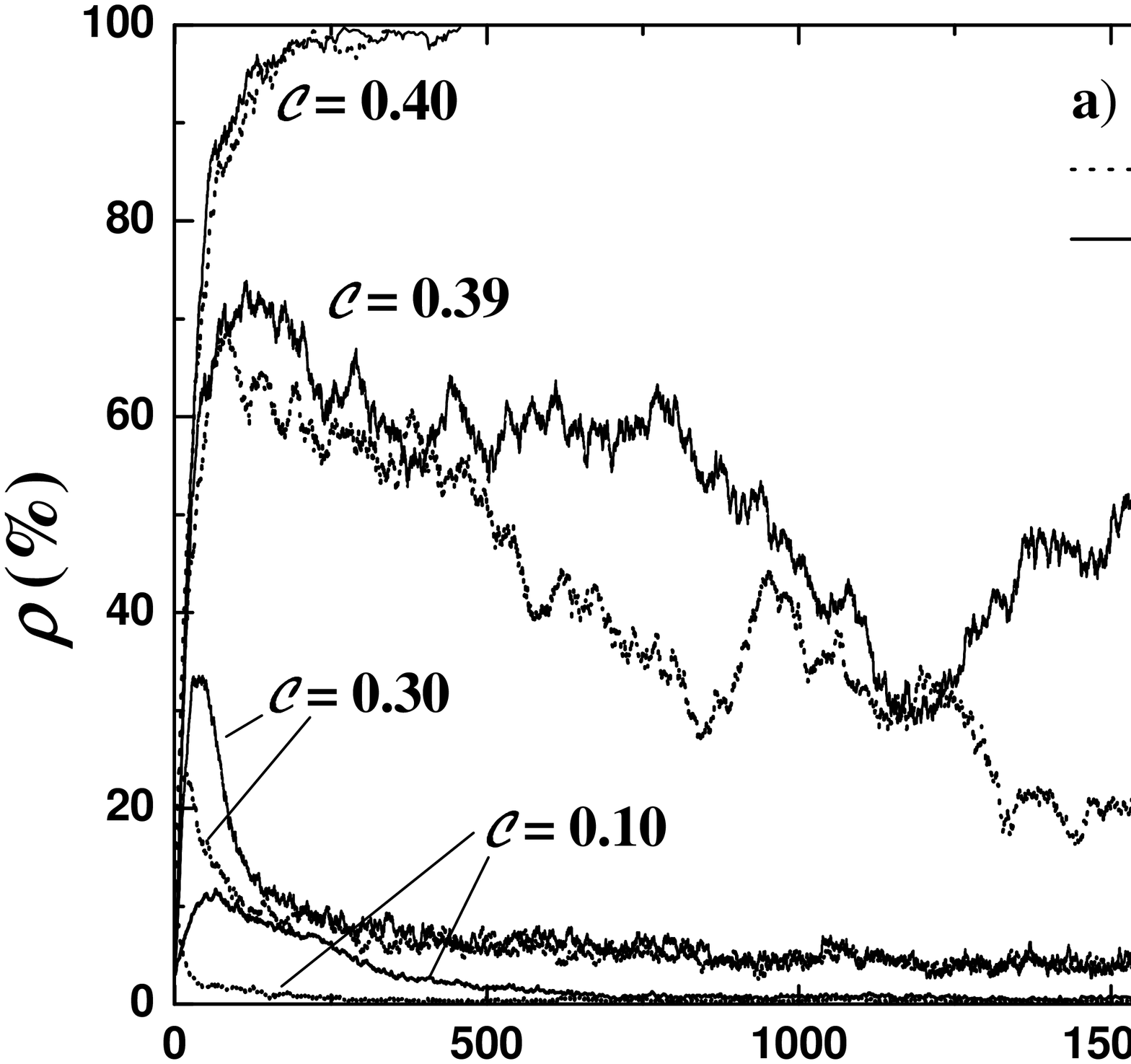,width=80mm} \psfig{file=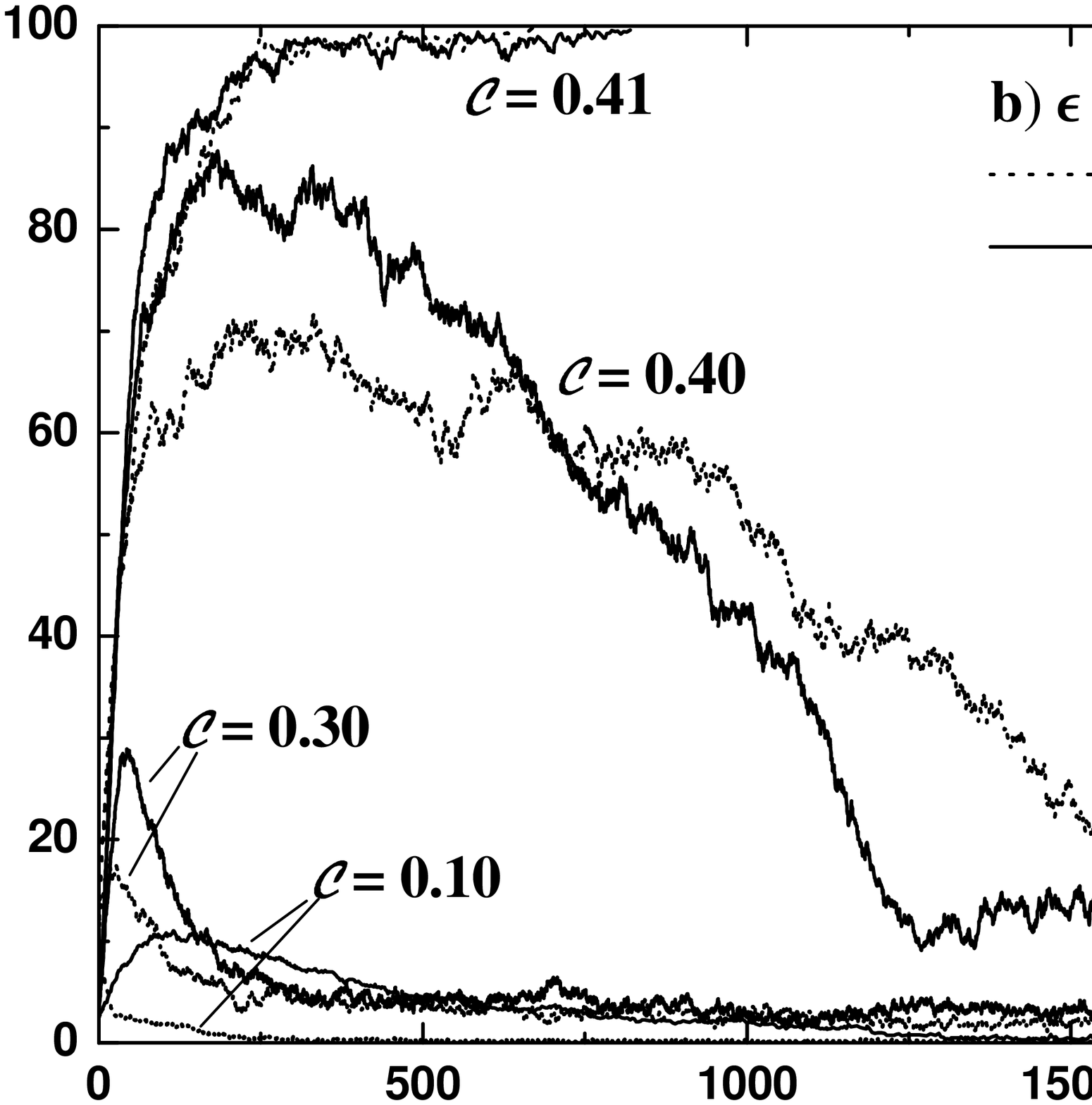,width=80mm}}
\centerline{\psfig{file=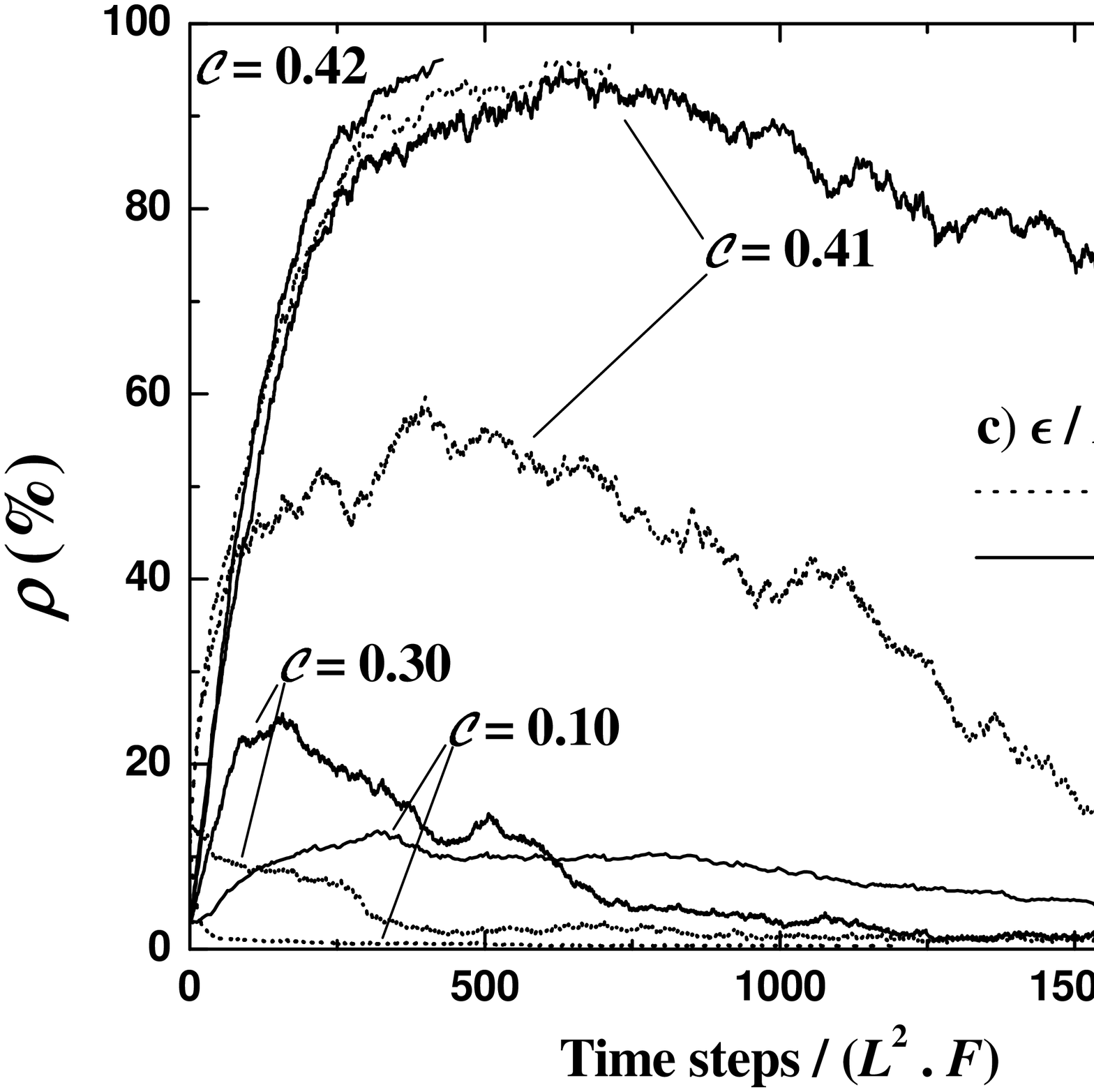,width=80mm}}
\vspace{-2cm}
\caption{Percent of the VF information in the lattice as a function of time for different values
of the ratio $\epsilon/F$ and the confidence $\cc$. In straight lines is used $q=34$ while in
dotted lines it is used $q=8$.}
\label{fig4}
\end{figure*}

Figure \ref{fig3} shows the calculations of $<S_{max}>$ as a function of $q$ at the absorbing
state. Each panel shows the result for a certain value of the ratio $\epsilon/F$ and different
values of the confidence $\cc$ in increasing order. The result using the Axelrod model without VF
is included for comparison with full square dots. In this case, the system reaches a monocultural
state at $q<q_c \approx 18$ and a multicultural state at $q>q_c$. In panel a) we set $\epsilon
= $ 1.0. It can be seen that when the value of the confidence $\cc$ is small ($\cc = $ 0.05), the
results are very close to those of Axelrod model. The monocultural states remain unchanged at
$q<q_c$ but the multicultural state is less ``robust'' and higher values of $<S_{max}>$ are
obtained. For increasing values of $\cc$, higher values of $<S_{max}>$ for $q>q_c$ are obtained, as
seen with $\cc = $ 0.10 (hence the number of different domains in the multicultural state are
smaller) and finally, at $\cc = $ 0.15, the multicultural states vanish and the system remains in
monocultural states for all values of $q$. Then, it can be concluded that the increasing value of
the confidence induces an homogenization of the cultural information that the system has, even at those
values of $q$ where the system reaches multicultural states when there is no VF. Then,
multicultural states are unstable for increasing values of the confidence $\cc$ at this value of
the effective trait $\epsilon$.

In Fig. \ref{fig3} b), c) and d) it is calculated the $<S_{max}>$ for smaller values of $\epsilon$.
It can be seen that multicultural states at $q>q_c$ are again obtained for low values of $\cc$ but
now higher values are needed for the confidence to produce a cultural homogenization ($<S_{max}
\approx 1>$). This can be seen when comparing the results for $\cc = $ 0.15 in Fig. \ref{fig3} a)
with $\epsilon=1$ and in Fig. \ref{fig3} b) with $\epsilon = $ 0.5. In Fig. \ref{fig3} c) it can
be seen that at a very small value of $\epsilon$ ($\epsilon = $ 0.1) the system has a
multicultural state for $q>q_c$ with small increases of $<S_{max}>$ for increasing $\cc$, and even
at $\epsilon = $ 0.01 (Fig. \ref{fig3} d)) the transition from a monoculture to a multiculture at
$q \approx 18$ is independent of the confidence for values between 0.0 and 0.2.

Nevertheless, care has to be taken when analyzing the information of the induced monoculture at
$q>q_c$ when the VF is present (as the case $\cc = $ 0.15 in Fig. \ref{fig3} a) and $\cc = $ 0.20
in Fig. \ref{fig3} b)). It is interesting to know whether the greatest domain in the absorbing
state characterized by $<S_{max}>$ has, or not, the information possesses by the VF in the
corresponding nominal features. This information is indicated in Fig. \ref{fig3}, where all the
full dots show absorbing states where the corresponding greatest domain does not possess the
information of the VF in any of its features. That is, the overlap between VF and the cultural
state of the largest domain is zero. Only at those absorbing states indicated by full stars ($\cc =
$ 1.00 in all panels and $\cc=$ 0.50 on Fig. \ref{fig3} d)) the corresponding biggest domain fully
shares the information at nominal features of the VF. Then, it is obtained the interesting result
that at low confidence values $\cc$, the culture homogenization induced by the VF results in a
negation or cancellation by the agents of the lattice of the information possesses by an external media. Here we
call this phenomenon {\it negative publicity} effect, and it represents the process occurring in
society when a group or different groups of people gather together, physically or intellectually, against an external
action they consider misconceived.

In order to study in more detail how the information of the VF spreads (or not) into the society,
we have defined the parameter $\rho$ which gives the percent of the total amount of the information
that agents share in their nominal features with the VF as a function of time. It is given by
\be
\rho = 100 \times \frac{1}{L^2 F}\sum_{i=1}^{L^2} \sum_{f=1}^F \delta_{\sigma_{i_f}, \sigma_{\m_f}}
\ee
and it is shown in Fig. \ref{fig4}. The calculation has been done for different values of the ratio
$\epsilon/F$ and different values of the confidence $\cc$ taking in consideration only one initial
state. In straight lines it is considered $q=34$ while in dotted lines $q=8$. When the dynamics
starts from random initial conditions, the information possessed by the VF is already present in
the lattice and shared by some agents. This anisotropy gives rise to the strong increase of its
percent at the beginning but if the confidence value is low, the information is avoided by agents
when traits are copied and it percent decreases with time after some maximum is achieved. This can
be seen for $\cc$ equal or below 0.39, 0.40 and 0.41 in Fig. \ref{fig4} a), b) and c) respectively.
It is necessary a higher value of $\cc$ for drive the system to a monocultural state with all
agents aligned with the VF information. On the other hand, if the confidence value is high enough,
the percentage of the VF information increases continuously until the absorbing state is reached. This
can be seen for $\cc = $ 0.40, 0.41 and 0.42 in Fig. \ref{fig4} a), b) and c) respectively. At the
same time, there is a sharp discontinuity between the regions of confidence values where the system
remains monoculture and multiculture, as can be seen between $\cc = $ 0.39 and 0.40 in Fig.
\ref{fig4} a), between $\cc = $ 0.40 and 0.41 in Fig. \ref{fig4} b) and between $\cc = $ 0.41 and
0.42 in Fig. \ref{fig4} c). For high values of $\cc$ the system reaches the monocultural state in
only few time steps, while for intermediate values of $\cc$ the value of $\rho$ changes slowly, at
least until $2000 L^2 F$ time steps where the simulation was artificially aborted. At those values
of $\cc$ the system seems to be in a quasi-stationary state and no absorbing state was found in
these regions. For enough low values of $\cc$, as those shown in Fig. \ref{fig3}, the system
reaches the multicultural state and there is almost no VF information on the lattice, as seen for
$\cc=0.10$ in Fig. \ref{fig4} in the three panels.

Finally, it is obtained that there are not fundamental differences between the behavior of the system
with $q=8$ and $q=34$ for higher values of $\epsilon$, as seen in Fig. \ref{fig4} a) and b).
Otherwise, in Fig. \ref{fig4} c) it is reported an important difference in the dynamics for $q=8$
and 34 in the region of confidence values where the system remains multicultural. This behavior is
connected with the strong decrease of $<S_{max}>$ in Fig. \ref{fig3} c) and d) for $\cc = $ 1.00
and for $\cc = $ 0.50 in Fig. \ref{fig3} d). In these cases, the high value of the confidence
quickly drives the system toward the absorbing state because of the strong assimilation of the VF
information by those agents that already share with it, at the beginning, the nominal traits. Then,
the system quickly breaks down into domains with null overlap.

\section{\label{conclusiones} Conclusions.}

It is developed here a new model for the inclusion of an external vector field in the Axelrod model
at zero temperature to describe the effects of the mass media on a social system. The clever design
of publicity which allows the mass media to have influence over the whole society was included as a
non-zero {\it extended} probability of traits being copied. This important effect is related with a parameter $\epsilon$ which can be interpreted as an extra effective feature or features the mass media could have with all agents in society, beyond the nominal features. This effective feature(s) is(are) used by the VF to reinforce the frequency of certain information already present on the society or to introduce a new one.

It is also included in the model a {\it confidence} value of the information possessed by the VF.
It is modeled as a probability of copying/deleting VF information which represents the criteria a person (or a group of persons) has (have) about what the mass media is proposing to the society. For very low values of this confidence, the
dynamics recovers the Axelrod model with no external field, but an increasing value produces an
homogenization on the society which would be multicultural without the external influence. This
cultural homogenization is lined up against the acting influence of the VF with a zero overlap with
the VF information. We have called here {\it negative publicity} to this effect. It simulates the
behavior of people in society who gathers together against the external information they estimate wrong or incorrect. For large values of the confidence, the system reaches a quasi-stationary state with
only slow changes of the amount of VF information in the society and no absorbing state was found
for the amount of time steps tested. At higher enough values of the confidence, it is obtained a
completed homogenized lattice lined up with the VF information, a situation that could be the
purpose of mass media, politic party, etc. These results are qualitatively in agreement with the
intuitive and realistic idea that with enough bombardment of the mass media information, that is accepted as valid and
trusted, there will be a strong induced culture homogenization in the society. In other words, when people
assume this information as personal the cultural differences tend to disappear.

We also studied the dependence with the effective probability and it was found that at small enough
values, the system is independent of the confidence value reproducing the results obtained in the
Axelrod model. For increasing value of the effective probability, lower values of the confidence
are needed to achieve monocultural systems lined up with the VF information.

It is important to mention that the confidence parameter $\epsilon$ can be experimental measured
and, therefore, it could advance the results expected on a society when designing publicity.

\section*{Acknowledgements}

The authors thank UNAM and CONACyT for partial financial support through Grants IN-114208 and
45835-F respectively. AHR also thanks psychologist Gezabel Guzm\'an for helpful suggestions and
enlightening discussions.

\end{document}